\documentclass[oneside, a4paper]{amsart}

\usepackage{graphicx}

\newcommand{\epsi}{{\varepsilon}}
\newcommand{\I}{{\rm i}}
\newcommand{\N}{\mathbb{N}}
\newcommand{\R}{\mathbb{R}}

\newcommand{\D}{{\rm d}}
\newcommand{\E}{{\rm e}}
\newcommand{\Hi}{\mathcal{H}}
\newcommand{\Or}{\mathcal{O}}

\newtheorem{theorem}{Theorem}[section]
\theoremstyle{definition}
\newtheorem{definition}[theorem]{Definition}

\begin{document}

\title{Effective Hamiltonians for thin Dirichlet tubes with varying cross-section}

\author{J. Lampart}
\address{Mathematisches Institut, Universit\"at  T\"ubingen\\
72076  T\"ubingen, Germany}

\author{S. Teufel$^*$}
\email{$^*$stefan.teufel@uni-tuebingen.de}
\urladdr{$^*$www.maphy.uni-tuebingen.de/members/stte}

\author{J. Wachsmuth}

\begin{abstract}
We show how to translate recent results on effective Hamiltonians for quantum systems constrained to a submanifold by a sharply peaked potential to quantum systems on thin Dirichlet tubes.   While the structure of the problem and the form of the  effective Hamiltonian stays the same, the difficulties in the proofs are   different.
\end{abstract}


\maketitle

\vspace{.8cm}

The question whether a Schr\"odinger Hamiltonian, which localizes states close to a submanifold of the configuration space by large forces, may be replaced by an effective operator on the submanifold is studied extensively and in various different settings in the literature (see e.g.\ \cite{FH01,Mi01,DE95,K08,DT04,BDT04,FS08,WT09,WT10}).

It is well-known that restricting the classical Hamiltonian system to the submanifold and then using Dirac's approach  to quantizing constrained Hamiltonian systems \cite{Di64} is too restricted. For there are lots of cases where the extrinsic curvature of the submanifold, which never shows up in Dirac's approach, plays a role. 
Therefore two other approaches have been investigated: 
\begin{itemize}
\item \emph{Soft constraints:} A rapidly increasing potential is used to localize solutions close to the submanifold (see \cite{FH01,Mi01} and references therein).
\item \emph{Hard constraint:} the localization is achieved via Dirichlet boundary conditions on a thin tube centered around the submanifold (see the reviews \cite{DE95,K08}).
\end{itemize}
If the potential or the tube's cross-section depend on the point on the submanifold, the constraint is called \emph{varying}. First results for such constraints were given in \cite{DT04,BDT04,FS08}.

Recently two of the authors have deduced effective Hamiltonians for the case of a varying soft constraint in arbitrary (co-)dimension (see \cite{WT09,WT10}). Here we explain how these results may also be obtained for the case of a varying hard constraint. 

\section{The Setting}
Let $(\mathcal{A},G)$ be a  Riemannian manifold  of dimension $d+k$ and  $\mathcal{C}$ a submanifold of dimension $d$ without boundary, which is equipped with the induced metric $g=G|_{T\mathcal{C}}$. 
We assume that there is a non-self-intersecting tube $\mathcal{B}_\delta$ of radius $\delta>0$ around $\mathcal{C}$. 

If $\mathcal{C}$ is compact, such a tube $\mathcal{B}_\delta$ always exists and is compact itself. For the sake of a simple presentation we will focus on the latter case  in the following and only shortly comment on the necessary adjustments in the case of a non-compact $\mathcal{C}$.

Since there is a canonical diffeomorphism $\Phi$ from $B_\delta$ into the normal bundle $\pi:N\mathcal{C}\to\mathcal{C}$, we can scale any subset of $B_\delta$ in the normal direction via
\[
D_\epsi: N\mathcal{C} \to N\mathcal{C}\,,\quad  (q,N) \mapsto (q,\epsi N)\,.
\]
Let $\Omega\subset B_\delta$ be an open subset with smooth boundary such that the cross-sections
\[
\Omega(q) := \Phi(\overline{\Omega})\cap N_q\,\mathcal{C}
\]
are all diffeomorphic, compact, and connected. Then $\pi:\Omega\to\mathcal{C}$ has the structure of a fiber bundle compatible with the one of $N\mathcal{C}$. We assume that this bundle has smooth local trivializations. In the case of a non-compact $\mathcal{C}$ one has to postulate the existence of a set of local trivializations whose derivatives satisfy global bounds in a suitable manner.
The $\epsi$-thin tube $\Omega^\epsi$ is now defined via
\[
\Omega^\epsi :=     \Phi^{-1} D_\epsi \Phi\,\Omega\,.
\]
Our goal is to approximate the spectrum of and the unitary group generated by
\[
 H^\epsi := -{\epsi^2} \Delta_G\quad \mbox{ on}\quad L^2(\Omega^\epsi,\mu_G)
 \]
with Dirichlet boundary conditions by using  an \emph{effective Schr\"odinger operator} $H_{\rm eff}^\epsi $ on $L^2(\mathcal{C})$. Here $\Delta_G$ is the Laplace--Beltrami operator associated with $G$ and the factor $\epsi^2$ has been put in for convenience because otherwise the spectrum of $H^\epsi$ would diverge in the limit $\epsi\to0$.
 $H^\epsi$ is obviously unitarily equivalent to the operator 
 \[
 - {\epsi^2} \Delta_{\Phi_*G}\quad\mbox{on}\quad L^2( \Phi(\Omega^\epsi), \mu_{\Phi_*G})
 \]
 with Dirichlet boundary conditions. We will identify the two operators in the following without making the diffeomorphism $\Phi$ explicit anymore.
 
\section{Basic Ideas} 
Consider the vector bundle $\mathcal{E}_{\rm f}:=\{(q,\varphi)\,|\,q\in\mathcal{C},\,\varphi\in C^\infty(\Omega(q))\}$ over $\mathcal{C}$, where the fibers $\Omega(q)$ of the bundle $\Omega$ are replaced with $C^\infty(\Omega(q))$ and the bundle structure of $\Omega$ is lifted by using the composition with the local trivializations of $\Omega$ as the new trivializations.
Via the normal connection $\nabla^\perp$ on $N\mathcal{C}$, which is induced by $G$,  every vector $\tau\in T_q\mathcal{C}$ tangent to $\mathcal{C}$ can be lifted into the tangent spaces $T_{q,n}\Omega$ of the corresponding fiber. The derivative of sections of $\mathcal{E}_{\rm f}$ into the direction of the lift defines the so-called \emph{horizontal connection} $\nabla^{\rm h}$ on $\mathcal{E}_{\rm f}$ (see~\cite{WT09}).
The associated Laplacian $\Delta_{\rm h}$ coincides with the Laplace--Beltrami operator $\Delta_g$ on $\mathcal{C}$ for functions that are constant on the fibers.
 
As in \cite{WT09,WT10} the basic idea is that after a measure transformation and rescaling the normal coordinates $n=N/\epsi$ the Hamiltonian $H^\epsi$ may be split as
\[
H^{\epsi} = -{\epsi^2}\Delta_{\rm h} - \Delta_n +  \Or({\epsi})\,.
\]
This suggests to define for each  $q\in \mathcal{C}$ the local  \emph{fiber Hamiltonian}
\[
H_{\rm f} (q) := -\Delta_n \quad \mbox{ on } \quad   L^2(\Omega(q),\D\lambda)
\]
with Dirichlet boundary conditions. Here $\D\lambda$ is the Lebesgue measure induced from $N_q\mathcal{C}\simeq\R^k$. 
Since each fiber $\Omega(q)$ is compact, the spectrum of $H_{\rm f}(q)$ is discrete for all $q\in\mathcal{C}$. At some fixed $q$ we number the eigenvalues by $J\in\N_0$. Due to the smooth dependence of $\Omega(q)$ on $q$ this gives rise to continuous families of eigenvalues $E_J(q)$, so-called \emph{energy bands}. In general, these bands may cross.

 \begin{definition}
An energy band $E_J$ is called \emph{admissible}, if $E_J(q)$ is simple for all $q\in\mathcal{C}$ and the associated complex eigenspace bundle is trivializable, i.e., there is a global  section $\varphi_J$ of normalized eigenfunctions. In addition, if $\mathcal{C}$ is non-compact, $E_J$ has to satisfy a gap condition as in \cite{WT09}. 
\end{definition}
As is well-known from the theory of elliptic operators, the lowest eigenvalue $E_0(q)$ on the connected domain $\Omega(q)$ is simple  and $\varphi_0(q)$ can be chosen positive. So the lowest energy band $E_0$ is always admissible because the positivity of $\varphi_0$ ensures that it is a global section.
For  an admissible energy band $E_J$ the subspace
\[
\mathcal{P}_J := \left\{ \psi(x)\varphi_J(x,n)\,|\, \psi\in L^2(\mathcal{C},g)\right\}\subset L^2(\Omega,G)
\]
may be identified with $L^2(\mathcal{C},g)$ via the unitary operator
\[
U_J: \mathcal{P}_J \to L^2(\mathcal{C},g)\,,\quad \psi(x)\varphi_J(x,n) \mapsto \psi(x).
\]
$\mathcal{P}_J$ is approximately invariant under $H^\epsi$
because the associated projector $P_J$ satisfies
\[
[P_J,H^\epsi]=[P_J,-\epsi^2\Delta_{\rm h}]+\Or({\epsi})=\Or({\epsi})
\]
in $\mathcal{L}(D(H^\epsi),\Hi)$. However, we are interested in the way the spectrum and the unitary group are affected by the geometry and the global structure of $\mathcal{C}$. These effects are of order $\epsi^2$. Therefore we have to improve on the invariance of the subspaces.

\section{Results}
Fix $E_{\rm max}<\infty$. Via adiabatic perturbation theory it is possible to construct a projector $P^\epsi_J=P_J+{\epsi} P_J^1+{\epsi^2} P_J^2$ and a unitary $U^\epsi_J:\mathcal{P}^\epsi_J\to L^2(\mathcal{C})$ such that 
\begin{equation}\label{invariance}
[P^\epsi_J,H^\epsi]\chi_{(-\infty,E_{\rm max}]}(H^\epsi)=\Or({\epsi^3}),
\end{equation}
where $\chi_{(-\infty,E_{\rm max}]}$ is the characteristic function of $(-\infty,E_{\rm max}]$.
The construction of $P^\epsi_J$ is quite similar to the one in \cite{WT09}. We comment on the differences below. Here we could in principle continue the construction to obtain a projector which is invariant up to errors of order $\epsi^N$ for any $N\in\N$. 

Now we reformulate the main result from \cite{WT09} for the case of thin Dirichlet tubes. Here we use the index formalism including the convention that  one sums over repeated indices. Moreover, we use latin indices $i,j,..$ running from $1$ to $d$ for coordinates on $\mathcal{C}$, greek indices $\alpha,\beta,\dots$ running from $d+1$ to $d+k$ for the normal coordinates, and latin indices $a,b,..$ running from $1$ to $d+k$ for coordinates on $\Omega$.

\begin{theorem}\label{theo}
Let $E_J$ be an admissible energy band and $E_{\rm max}<\infty$. There are $C<\infty$ and $\epsi_0>0$ such that for all $\epsi<\epsi_0$ there exist a Riemannian metric $g_J^\epsi$ on~$\mathcal{C}$,  an orthogonal  projection $P_J^\epsi$, a unitary
$U_J^\epsi: P_J^\epsi L^2(\Omega,\mu_G) \to L^2(\mathcal{C},\mu_{g_J})$
and
\[
H_J^\epsi:= U_J^\epsi P_J^\epsi  H^\epsi P_J^\epsi U_J^{\epsi *}  \quad\mbox{with domain}\quad U_J^{\epsi}D(H^\epsi),
\]
which satisfy the following:
\begin{enumerate}
\item[(a)] {\bf Dynamics:} $H_J^{\epsi}$ is self-adjoint on $L^2(\mathcal{C},\mu_{g_J})$ and
\[
\left\| \, \left( \E^{-\I H^\epsi t } -  U_J^{\epsi *}\,\,\E^{-\I H_J^\epsi t }\,\, U_J^{\epsi}\right)\,P_J^\epsi\,\,\chi_{(-\infty,E_{\rm max}]}(H^\epsi)\right\|\leq C\,{\epsi^3 }|t|.
\]

\smallskip

\item[(b)] {\bf Spectrum:}
For all $(E^\epsi)$ with $\limsup_\epsi E^\epsi< E_{\rm max}$ one has
\begin{enumerate}  
\item[(i)] \quad $H_J^\epsi\,\psi^\epsi = E^\epsi\,\psi^\epsi$\quad $\Rightarrow$\quad $ \left\| \left( H^\epsi - E^\epsi\right)  U_J^{\epsi*} \psi^\epsi\right\| \leq C \,{\epsi^3}\, \| U_J^{\epsi*} \psi^\epsi \| $,\vspace{0.1cm}
\item[(ii)] \quad $ H^\epsi  \, \Psi^\epsi = E^\epsi  \,\Psi^\epsi$\quad $\Rightarrow$\quad $
\| ( H^\epsi_J-E^\epsi)\,  U_J^\epsi P_J^\epsi  \Psi^\epsi \| \leq C\,{\epsi^3}\, \|  \Psi^\epsi \|$.
\end{enumerate}

 \end{enumerate}

\medskip

For $\psi_1=\chi_{(-\infty,E_{\rm max}]}(-\epsi^2\Delta_g+E_J)\psi_1$ the effective Hamiltonian $H^\epsi_J$ is given by
\begin{eqnarray*}
\langle\psi_2|H^\epsi_J\psi_1\rangle_\mathcal{C} 
&=& \int_\mathcal{C} \Big(g_J^{\epsi\,ij}\,\overline{p^J_{\epsi\,i}\psi_2}\,p^J_{\epsi\,j}\psi_1
\,+\,\overline{\psi_2}E_J\psi_1\,-\,\epsi^2\,\overline{\psi_2}\,U_1^{\epsi\,*}R_{H_{\rm f}}(E_J)U_1^\epsi\,\psi_1 \\
&& \qquad\qquad\qquad\quad+\,\epsi^2\overline{\psi_2}\left(V_{\rm geom}+V_{{\rm BH}}+V_{{\rm amb}} \right)\psi_1 \Big)d\mu_{g_J^\epsi} \,+\,\Or(\epsi^3),
\end{eqnarray*}
where
\begin{eqnarray*}
g_J^{\epsi\,ij} &=& g^{ij}+ \epsi \,2{\rm II}^{ij}_\alpha \langle \varphi_J| n^\alpha \varphi_J\rangle_{\Omega(q)} +\epsi^2\overline{\mathcal{R}}^{i\ j\;}_{\,\alpha\;\beta}\big\langle\varphi_J\big|n^\alpha n^\beta\varphi_J\big\rangle_{\Omega(q)}\\
&& \ +\epsi^2\mathcal{W}_{\alpha l}^{i}g^{lm}\mathcal{W}_{\beta m}^{j}\big\langle\varphi_J\big|3n^\alpha n^\beta \varphi_J\big\rangle_{\Omega(q)},\vspace{0.5cm}\\
p^J_{\epsi\,j} &=& -\I \epsi \partial_{j} -\epsi \langle \varphi_J| \I  \nabla^{\rm h}_{j}\varphi_J\rangle_{\Omega(q)} -\epsi^2 \overline{\mathcal{R} }^{\ \ \gamma\;}_{j\alpha\;\beta}\langle\varphi_J|{\textstyle \frac{2}{3}}n^\alpha n^\beta\I\partial_\gamma\varphi_J\rangle_{\Omega(q)}\\
&& \,+\epsi^2\mathcal{W}_\alpha^{ji}\big\langle\,\varphi_J\,\big|\,2\,\big(n^\alpha-\langle\varphi_J|n^\alpha\varphi_J\rangle\big)
\I\nabla^{\rm h}_i\varphi_J\,\big\rangle_{\Omega(q)},\vspace{0.5cm}\\
R_{H_{\rm f}}(E_J) &=& (1-P_J)\big(H_{\rm f}-E_J\big)^{-1} (1-P_J),\vspace{0.5cm}\\
U_1^\epsi &=& 2g^{ij}\overline{\nabla^{\rm h}_i \varphi_J}\epsi\partial_{j}+n^\alpha\varphi_J\mathcal{W}^{ij}_\alpha\epsi^2\partial^2_{ij},\vspace{0.5cm}\\
V_{{\rm geom}} &=& -{\textstyle \frac{1}{4}}\eta^{\alpha}\eta_\alpha+{\textstyle \frac{1}{2}}\mathcal{R}_{\;\;ij}^{ij}-{\textstyle \frac{1}{6}}\big(\overline{\mathcal{R}}_{\;\;ab}^{ab}+\overline{\mathcal{R}}_{\;\;aj}^{aj}+\overline{\mathcal{R}}_{\;\;ij}^{ij}\big),\vspace{0.5cm}\\
V_{{\rm BH}} &=& g^{ij}\big\langle\nabla^{\rm h}_i\varphi_J\big|(1-P_J)\nabla^{\rm h}_J\varphi_J\big\rangle_{\Omega(q)},\vspace{0.5cm}\\
V_{{\rm amb}} &=& 
\overline{\mathcal{R}}^{\gamma\ \delta\;}_{\;\alpha\;\beta}\langle\partial_\gamma\varphi_J|{\textstyle \frac{1}{3}}n^\alpha n^\beta\partial_\delta\varphi_J\rangle_{\Omega(q)},
\end{eqnarray*}
with $\mathcal{W}$ the Weingarten mapping, $\eta$ the mean curvature vector, $\mathcal{R}$ and $\overline{\mathcal{R}}$ the Riemann tensors of $\mathcal{C}$ and $\mathcal{A}$ (see \cite{WT10} for definitions of all the geometric objects).
\end{theorem}

For a non-compact $\mathcal{C}$ additional bounds on the derivatives of $\varphi_J$ as in \cite{WT09} are required. There a detailed discussion of the effective Hamiltonian is provided, too. 

\smallskip

As soon as (\ref{invariance}) has been established, the proof of Theorem \ref{theo} goes exactly along the same lines as in \cite{WT09}. The strategy to obtain (\ref{invariance}) is also the same here, but the technical difficulties are different. The key facts that have to be derived are
\begin{eqnarray}
\| [-\epsi^2\Delta_{\rm h},P_J]\|_{\mathcal{L}(D(H_\epsi^{m+1}),D(H_\epsi^m))} &=& \Or(\epsi), \label{one}\\
\|[-\epsi^2\Delta_{\rm h},R_{H_{\rm f}}(E_J)]\|_{\mathcal{L}(D(H_\epsi^{m+l}),D(H_\epsi^m))} &=& \Or(\epsi) \label{two}
\end{eqnarray}
for some $l\in\N$ and all $m\in\N_0$.
In addition, in \cite{WT09} we had to make sure that the derivatives of $\varphi_J$ decay fast enough in the spatially infinite fibers and that their decay is not destroyed by application of energy cutoffs and resolvents, which is not necessary in the case considered here due to the boundary conditions. However, the boundary poses new problems in the proof of (\ref{one}) \& (\ref{two}). On the one hand, the volume of the fibers is varying so that $\nabla^{\rm h}$ is only metric on sections which satisfy the Dirichlet condition. On the other hand, application of $\nabla^{\rm h}$ destroys the Dirichlet condition. Therefore one cannot only stick to the differential operators but has to make use of the spectral representation, too.

Roughly speaking, (\ref{one}) means to show that all the derivatives of $\varphi_J$ are uniformly bounded, in particular at the boundary. This can be done by locally mapping $\Omega$ to the constant tube equipped with a suitable product metric and applying the procedures from \cite{WT09}. Here the smoothness of the trivializations of $\Omega$ enters.

For (\ref{two}) one makes use of the fact that the fibers are compact so that the resolvent may be written as $R_{H_{\rm f}}(E_J)=\sum_{I\neq J}P_I/(E_I-E_J)$. Then its derivatives may be controlled via some Weyl's law by choosing $l$ large enough.


%
%

\section{Discussion of the Results}

Due to Theorem \ref{theo} the spectrum of $H^\epsi$ is given, up to errors of order $\epsi^3$, by the spectra of $H_J^\epsi$ for $J\in\N_0$. With our approach it is possible to obtain not only the energies close to $\inf\sigma(H_J^\epsi)$ but also the excitations of order $1$. In this energy regime the leading part $-\epsi^2\Delta_g+E_J$ is a semiclassical operator, whose dynamics explores distances of order $1$ for times of order $\epsi^{-1}$. Therefore this is the relevant time scale, on which the global structure of an $\epsi$-independent $\mathcal{C}$ is seen. Theorem \ref{theo} allows to look at even much longer times.

The spectrum of $-\epsi^2\Delta_g+E_J$ is quite well-understood. 
We discuss here the role of the corrections in $H_J^\epsi$ for constant $E_J$ and an $E_J$ with one non-degenerate minimum on a compact $\mathcal{C}$. By using standard results from semiclassical analysis this discussion could be extended also to several degenarate minima.

\smallskip

\emph{1) $E_J$ is constant:} In this case the level spacing of $-\epsi^2\Delta_g+E_J$ is of order $\epsi^2$ close to $E_J$. So the low eigenvalues strongly depend both on the corrections in $p^J_\epsi$ and on the effective potentials $V_{\rm geom}$ and $V_J$ (see \cite{WT10} for an example with global effects). Since the kinetic energy is small for eigenvalues close to $E_J$, the corrections to $g$ and the off-band coupling $U_1^{\epsi\,*}R_{H_{\rm f}}(E_J)U_1^\epsi$ do not matter here.

They only become relevant for energies of order $1$ above $E_J$. Let $0\leq\alpha<2$. According to Weyl's law, the level spacing at energies of order $\epsi^\alpha$ above $E_J$ is of order $\epsi^{\alpha(1-d/2)+d}$. For $d=1,2$ this is always bigger than the approximation error of order~$\epsi^3$. For $d\geq3$ the approximation error is only smaller for $\alpha>(2d-6)/(d-2)$. Note that the minimal $\alpha$ is always strictly smaller than $2$. To fully resolve the spectrum for energies of order $1$ one would have to go to order $d$ in the construction of the super-adiabatic projector and the effective Hamiltonian. However, even in cases where the effective Hamiltonian is not precise enough
to resolve the small level spacing,
Theorem \ref{theo} still yields good control over the dynamics of states in this energy regime on the relevant time scales.

\smallskip

\emph{2) $E_J$ has one non-degenerate minimum:} 
Order $1$ above $\sup E_J$ the level spacing of $H_J^\epsi$ is again given by Weyl's law, i.e., only of order $\epsi^d$. 
So we are in the same situation as for energies of order $1$ above a constant $E_J$. Close to $\sup E_J$ no general statements can be made about the level spacing. Order $1$ below $\sup E_J$ the spectrum of $H_J^\epsi$ is dominated by $E_J$ resulting in a level spacing of order~$\epsi$. Thus the effective potentials, which are of order $\epsi^2$, may be ignored here. For energies of order $1$ above $\inf E_J$, however, the $\epsi$-corrections to the kinetic energy become relevant because the eigenfunctions oscillate on a scale of order $\epsi^{-1}$. 

We sketch the emerging picture for the case of a closed curve $\mathcal{C}$ of length~$L$:
\begin{figure}[h]
\begin{center}
\includegraphics[height= 5cm]{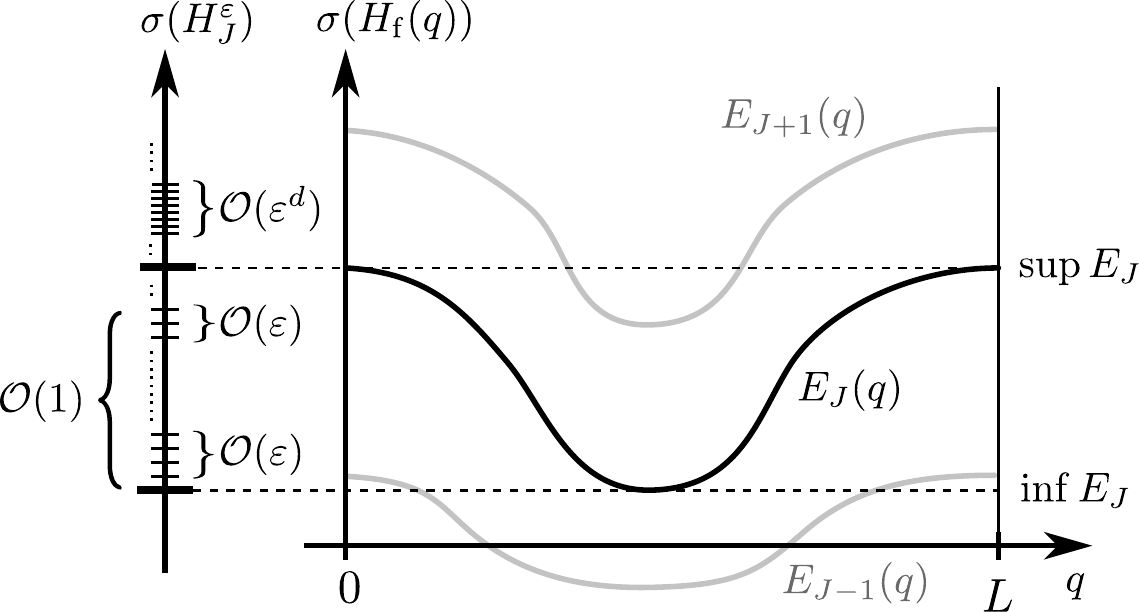}
\end{center}
\end{figure}

For energies above $\sup E_J$ the corresponding eigenfunctions are not localized but extended over the whole submanifold. Hence, global effects may occur here, too.


\section{Conclusions and Outlook}

We have derived an effective Hamiltonian for the problem of hard constraints in quantum mechanics that covers all the interesting energy scales.
Although the technical difficulties in the soft and the hard constraint approach differ, the results have a very similar structure. This is due to the fact that the wave function concentrates close to the submanifold.
In the future, we will investigate the Laplacian on thin Riemannian fiber bundles using our adiabatic techniques. To do so we split up the metric into a horizontal and a vertical part and scale only the latter by $\epsi$:
\[
G=G_{\rm h}+\epsi^2G_{\rm v}\,.
\]
This is related to the so-called adiabatic limit in global analysis (see e.g.\ \cite{MM90}). In this setting there is no concentration inside the fibers. Therefore an expansion of the metric must be replaced by an averaging procedure. As a consequence,  the effective Hamiltonian will have a somewhat different structure.


\begin{thebibliography}{9}




\bibitem{FH01} R.\ Froese, I.\ Herbst, {\it Realizing Holonomic Constraints in Classical and Quantum Mechanics}, Commun.\ Math.\ Phys.\ {\bf 220}, 489--535 (2001).

\bibitem{Mi01} K.\ A.\ Mitchell, {\it Gauge fields and extrapotentials in constrained quantum systems}, Phys.\ Rev.\ A {\bf 63}, 042112 (2001).

\bibitem{DE95} P.\ Duclos, P.\ Exner, {\it Curvature-induced bound states in quantum waveguides in two and three dimensions}, Rev.\ Math.\ Phys.\ {\bf 7}, 73--102 (1995).

\bibitem{K08} D.\ Krej$\check{{\rm c}}$i$\check{\rm r}$\'ik, {\it Twisting versus bending in quantum wave guides}, in Analysis on Graphs and its Applications, Proc.\ Sympos.\ Pure Math.\ {\bf 77}, Amer.\ Math.\ Soc., 617--636 (2008), see arXiv:0712.3371v2 [math-ph] for a corrected version. 

\bibitem{DT04} G.\ F.\ Dell'Antonio, L.\ Tenuta, {\it Semiclassical analysis of constrained quantum systems},  J.\ Phys.\ A {\bf 37}, 5605--5624 (2004).

\bibitem{BDT04} V.\ V.\ Belov, S.\ Yu.\ Dobrokhotov,  T.\ Ya.\ Tudorovskiy, {\it Asymptotic solutions of nonrelativistic equations of quantum mechanics in curved nanotubes}, Theo.\ Math.\ Phys.\   {\bf 141}, 1562--1592 (2004).


\bibitem{FS08} L.\ Friedlander, M.\ Solomyak, {\it On the spectrum of the Dirichlet Laplacian in a narrow infinite strip}, in Spectral theory of differential operators. M.\ S.\ Birman 80th anniversary collection, AMS Translations, Series 2, Advances in the Mathematical Sciences {\bf 225} (2008).


\bibitem{WT09} J.\ Wachsmuth, S.\ Teufel, {\em Effective Hamiltonians for Constrained Quantum Systems}, e-print arXiv:0907.0351v3.
 
\bibitem{WT10} J.\ Wachsmuth, S.\ Teufel, {\em Constrained Quantum Systems as an Adiabatic Problem}, Phys.\ Rev.\ A {\bf 82}, 022112 (2010).
 
\bibitem{Di64} P.\ A.\ M.\ Dirac, {\it Lectures on Quantum Mechanics}, Yeshiva Press (1964).

\bibitem{MM90} R.\ R.\ Mazzeo, R.\ B.\ Melrose, {\em The adiabatic limit, Hodge cohomology and Leray's spectral sequence for a fibration}, J.\ Diff.\ Geom.\ {\bf 31}, 185--213 (1990).
  \end{thebibliography}
\end{document}